\documentclass[aps,showpacs,amssymb,floatfix,pre,notitlepage]{revtex4-1}
\usepackage[dvips]{graphicx}
\usepackage[activeacute,english]{babel}
\usepackage{colordvi,amsmath,epsfig,color}

\newcommand{\be}{\begin{equation}}
\newcommand{\ee}{\end{equation}}

\begin{document}

\title{Universal reference state in a driven homogeneous granular gas}
\author{Mar\'ia Isabel Garc\'ia de Soria}
\affiliation{F\'{\i}sica Te\'{o}rica, Universidad de Sevilla,
Apartado de Correos 1065, E-41080, Sevilla, Spain}
\author{Pablo Maynar}
\affiliation{F\'{\i}sica Te\'{o}rica, Universidad de Sevilla,
Apartado de Correos 1065, E-41080, Sevilla, Spain}
\author{Emmanuel Trizac}
\affiliation{Laboratoire de Physique Th\'eorique et Mod\`eles Statistiques, 
UMR CNRS 8626, Universit\'e Paris-Sud, 91405 Orsay, France}

\begin{abstract}
We study the dynamics of a homogeneous granular gas heated by a stochastic 
thermostat, in the low density limit. It is found that, before reaching the 
stationary regime, the system quickly ``forgets'' the initial condition and then evolves  
through a universal state that does not only depend on the
dimensionless velocity, but also on the instantaneous temperature,
suitably renormalized by its steady state value. We find 
excellent agreement between the theoretical predictions at Boltzmann equation level
for the one-particle 
distribution function, and Direct Monte Carlo simulations. 
We conclude that at variance with the homogeneous cooling phenomenology,
the velocity statistics should not be envisioned as a single-parameter,
but as a two-parameter scaling form, keeping track of the distance to
stationarity.

\end{abstract}
\maketitle

\section{Introduction}
\label{sec:intro}

A granular gas may be viewed as a collection of macroscopic particles
undergoing dissipative collisions. This very ingredient --inelasticity--
gives rise to a rich phenomenology \cite{d00,g03,BP04,at06}, the understanding
of which requires statistical mechanics tools: Kinetic theory 
has proven to be powerful at a microscopic/mesoscopic level of description, while
at a macroscopic scale, hydrodynamic equations have been derived \cite{g03} and put to the test. 
Yet, the relevance and consistency of a hydrodynamic 
framework, which is a central question, is still elusive \cite{k99,g01,d01}. 

Due to collisional dissipation, the granular temperature, defined as the 
variance of velocity fluctuations, 
decays monotonically in time in an isolated granular system \cite{h83}. In the 
fast-flow regime, 
it has been shown numerically that for a 
wide class of initial conditions, the system reaches a homogeneous state in 
which all the time dependence of the one-particle distribution function is encoded in the 
temperature. 
This is the so-called \emph{homogeneous cooling state} (HCS) which has 
been widely studied in the literature \cite{gs95, vne98}. In such a 
homogeneous situation, the dynamics involves two time-scales: the kinetic 
or fast scale --a few collisions per particle-- 
in which the scaling regime has not been reached yet and where 
the ``microscopic'' excitations relax, and the following hydrodynamic 
or slow scale in which the memory of initial conditions has been lost, 
and the velocity distribution evolves through 
the granular temperature \cite{bdr03}.
Considering non 
homogeneous states, this separation of time scales opens the possibility of a
hydrodynamic or coarse-grained description 
in terms of the density, velocity and temperature field, 
and the HCS then plays the role of the reference state when the 
Chapmann-Enskog method is applied \cite{bdks98}. 

On the other hand, several studies and experiments in granular matter deal
with stationary 
states, which are reached under the action of some energy driving,
often realized by a moving boundary, or by an 
interstitial medium that acts as a thermostat, see e.g. 
\cite{PEU02,L09,PGG12}. In all these cases, the energy 
injected compensates for the energy lost in collisions. From a theoretical point of 
view, a minimalistic approach is to consider 
the system as driven by some random energy source, which can be implemented in different 
ways \cite{clh00}. 
For the hard particle model, 
one of the most used homogeneous heating is the so-called stochastic 
thermostat, which consists in a white noise force acting on each grain 
\cite{wm96,vne98,vnetp99,MS00,P02,gm02,V06,V06bis,M09,M09bis,VL11,plmv99}. 
In the low density limit, the distribution function 
of the homogeneous state 
has been characterized \cite{vne98}. Hydrodynamic equations have been 
derived via the Chapmann-Enskog expansion \cite{gm02} and fluctuating 
hydrodynamics have been put forward in order to understand the large 
scale structure found in the stationary state \cite{vnetp99}, 
or fluctuations of global quantities \cite{M09bis}. We stress
that in all the studies pertaining to the hydrodynamics of a system heated by the 
stochastic thermostat \cite{vnetp99,gm02}, 
the stationary state played the role of ``reference'' state, 
as the HCS happens to be in 
the undriven case. It was therefore assumed that a one parameter scaling holds
for the velocity probability distribution.
The objective in this work is to analyze this point 
critically in the homogeneous case at the level of the Boltzmann equation. 
We will study, for arbitrary initial 
conditions, the type of state the system evolves into in a kinetic time scale. 
Surprisingly, we find that a universal state is reached in a kinetic scale 
--universal in the sense that it is independent of the initial conditions-- but that 
depends on the quotient between the instantaneous temperature and its 
stationary value. 

The outline is as follows. Section \ref{sec:scaling} opens with a definition 
of the model, and a summary of relevant previously known results. The key question 
addressed lies in the scaling form of the velocity distribution close to
the steady state. Does it depend only on the suitably reduced velocity variable,
as is the case in the HCS, or is another parameter relevant, that would encode the
distance to stationarity ? We will argue in section \ref{ssec:oneortwo} that a 
single parameter scaling form is inconsistent. We shall then show in
sections \ref{ssec:numerics} and 
\ref{ssec:analytical} that a consistent two parameter scaling form can be identified.
Its properties will be characterized by complementary numerical and analytical
tools. Conclusions and perspectives will finally be discussed in section 
\ref{sec:concl}.

\section{Scaling form of the velocity distribution function}
\label{sec:scaling}

The system of interest is a dilute gas of $N$ smooth inelastic hard particles
of mass $m$ and diameter $\sigma$, which collide inelastically with a 
coefficient of normal restitution $\alpha$ independent of the relative 
velocity \cite{BP04}. If at time $t$ there is a binary encounter between particles 
$i$ and $j$, having velocities $\mathbf{V}_i(t)$ and $\mathbf{V}_j(t)$ 
respectively, the post-collisional velocities $\mathbf{V}_i'(t)$ and 
$\mathbf{V}_j'(t)$ are
\begin{eqnarray}\label{collisionRule}
\mathbf{V}_i'&=&\mathbf{V}_i-\frac{1+\alpha}{2}
(\hat{\boldsymbol{\sigma}}\cdot\mathbf{V}_{ij})\hat{\boldsymbol{\sigma}},
\nonumber\\
\mathbf{V}_j'&=&\mathbf{V}_j+\frac{1+\alpha}{2}
(\hat{\boldsymbol{\sigma}}\cdot\mathbf{V}_{ij})\hat{\boldsymbol{\sigma}},
\end{eqnarray}
where $\mathbf{V}_{ij}\equiv\mathbf{V}_i-\mathbf{V}_j$ is the relative
velocity and $\hat{\boldsymbol{\sigma}}$ is the unit vector pointing from the
center of particle $j$ to the center of particle $i$ at contact. Between 
collisions, the system is heated uniformly by a white noise acting independently on 
each grain \cite{vne98,vnetp99,P02,gm02,V06,M09,M09bis}
so that the one-particle velocity distribution, $f(\mathbf{r},\mathbf{v},t)$, then obeys 
the 
Boltzmann-Fokker-Planck equation \cite{vne98,vk92}. For a homogeneous system 
this equation reads
\be\label{ec.b}
\frac{\partial}{\partial t}f(\mathbf{v}_1,t)=
\sigma^{d-1}\int d\mathbf{v}_2\bar{T}_0(\mathbf{v}_1,\mathbf{v}_2)
f(\mathbf{v}_1,t)f(\mathbf{v}_2,t)
+\frac{\xi_0^2}{2}\frac{\partial^2}{\partial\mathbf{v}_1^2}f(\mathbf{v}_1,t),
\ee
where $d$ is the dimension of space, 
$\xi_0$ measures the noise strength,  
and $\bar{T}_0$ is the binary 
collision operator 
\be
\bar{T}_0(\mathbf{v}_1,\mathbf{v}_2)=\int d\hat{\boldsymbol{\sigma}}
\Theta(\mathbf{v}_{12}\cdot\hat{\boldsymbol{\sigma}})
(\mathbf{v}_{12}\cdot\hat{\boldsymbol{\sigma}})(\alpha^{-2}b_{\sigma}^{-1}-1).
\ee
Here we have introduced the operator $b_{\sigma}^{-1}$ which replaces the 
velocities $\mathbf{v}_1$ and $\mathbf{v}_2$ by the precollisional ones 
$\mathbf{v}_1^*$ and $\mathbf{v}_2^*$ given by 
\begin{eqnarray}
\mathbf{v}_1^*&=&\mathbf{v}_1-\frac{1+\alpha}{2\alpha}
(\hat{\boldsymbol{\sigma}}\cdot\mathbf{v}_{12})\hat{\boldsymbol{\sigma}},
\nonumber\\
\mathbf{v}_2^*&=&\mathbf{v}_2+\frac{1+\alpha}{2\alpha}
(\hat{\boldsymbol{\sigma}}\cdot\mathbf{v}_{12})\hat{\boldsymbol{\sigma}}.
\end{eqnarray}

\subsection{One-parameter scaling or beyond ?}
\label{ssec:oneortwo}

It is an observation from numerical simulations, 
that for a wide class of initial conditions the system 
reaches a stationary state \cite{vne98,vnetp99,P02}.  
Assuming that total momentum is zero, i.e. 
$\int d\mathbf{v}\mathbf{v}f(\mathbf{v},0)=\mathbf{0}$, the state is 
characterized by an isotropic stationary 
distribution, $f_s(v)$. Let us define the scaled distribution 
function, $\chi_s$, by
\be
f_s(v)=\frac{n}{v_s^d}\chi_s(c), \qquad 
\mathbf{c}=\frac{\mathbf{v}}{v_s}, 
\ee
where $n$ is the density, $v_s\equiv \sqrt{2T_s/m}$ is the thermal velocity and 
$T_s$ is the stationary temperature, 
$\frac{d}{2}nT_s=\int d\mathbf{v}\frac{1}{2}mv^2f_s(\mathbf{v})$. 
As $\chi_s$ is rather close to 
a Maxwellian distribution, a reasonable strategy is to perform 
an expansion in Sonine 
polynomials \cite{resibois}. In the 
so-called first Sonine approximation, the steady state function then reads \cite{vne98}
\be
\chi_s(c)\approx\chi_M(c)[1+a_2^sS_2(c^2)], 
\ee
where $\chi_M$ is the Maxwellian distribution with unit temperature, 
$S_2(c^2)=\frac{d(d+2)}{8}-\frac{d+2}{2}c^2+\frac{1}{2}c^4$, is the second 
Sonine polynomial, and $a_2^s$ is the kurtosis of the distribution. Within this 
approximation, the distribution function can be calculated, with the result \cite{vne98}
\begin{equation}
a_2^s(\alpha)=\frac{16(1-\alpha)(1-2\alpha^2)}
{73+56d-24d\alpha-105\alpha+30(1-\alpha)\alpha^2},
\label{eq:a2s}
\end{equation}
and a stationary temperature 
\begin{equation}
T_s=m\left[\frac{d\Gamma(d/2)\xi_o^2}
{2\pi^{\frac{d-1}{2}}(1-\alpha^2)n\sigma^{d-1}}\right]^{2/3}. 
\end{equation}

Now, let us consider an initial condition with a temperature that differs 
appreciably from the stationary temperature (we also assume that total 
momentum is zero, its precise value being immaterial). 
It is clear that the system will reach the stationary 
state in a hydrodynamic scale. The ensuing question is two-pronged.
First of all, is the dynamics compatible with a universal scaling 
form --once the memory of initial condition is lost-- that would
provide a consistent solution to the Boltzmann equation,
or is memory only washed out strictly speaking at the steady state point? 
Second, assuming such a scaling regime exists in some vicinity
of the steady state, what is the minimal number of parameters
required for its description? By analogy with unforced (HCS) phenomenology,
a single parameter scaling might be anticipated:
\begin{equation}
 f(\mathbf{v},t)=\frac{n}{v_o^d(t)}\chi_s(c) \quad 
 \hbox{where} \quad \mathbf{c}\equiv\mathbf{v}/v_0(t) \quad 
 \hbox{and} \quad v_0(t)\equiv\sqrt{2T(t)/m}
 \label{eq:naive}
\end{equation}
is defined from the instantaneous temperature
$d nT(t)=\int d\mathbf{v} m v^2f(\mathbf{v},t)$. 
We note that this scaling property 
holds for the Gaussian thermostat as well \cite{MS00}, where the particles  
are accelerated between 
collisions by a force proportional to its own velocity 
\cite{gT}. Moreover, in the stochastic thermostat case, 
the one parameter scaling (\ref{eq:naive}) 
was implicitly assumed, and it seemed to be yield 
reasonable predictions at least close to the stationary state, see \cite{vnetp99,gm02}. 
Nevertheless, when the form (\ref{eq:naive})
is inserted in the Boltzmann equation, Eq. (\ref{ec.b}), we 
obtain
\begin{equation}\label{in.ec}
\frac{\xi_0^2}{2v_0^3(t)}\frac{\partial^2}{\partial\mathbf{c}_1^2}\chi_s(c_1)
+\frac{1}{v_0^2(t)}\frac{dv_0(t)}{dt}\frac{\partial}{\partial\mathbf{c}_1}
\cdot [\mathbf{c}_1\chi_s(c_1)]
=-n\sigma^{d-1}\int d\mathbf{c}_2\bar{T}_0(\mathbf{c}_1,\mathbf{c}_2)
\chi_s(c_1)\chi_s(c_2), 
\end{equation}
which is inconsistent: The left hand side depends on time while the 
right hand side does not. We conclude that such a 
solution should be ruled out, except when stationarity is reached and $v_0=v_s$.
The hope is to capture the post-kinetic time dependence of 
$ f(\mathbf{v},t)$, through a more involved functional form,
that would be free of the above inconsistency.

As is customary, we shall seek for a normal solution \cite{resibois},
which in the present homogeneous case means that 
$f(\mathbf{v},t)$ should only depend on time via the instantaneous granular
temperature $T(t)$. In conjunction with dimensional analysis,
this leads to a function that should only depend on $\mathbf{c}$
and $T(t)/T_s = v_0^2/v_s^2$, which we write as
\be
f(\mathbf{v},t)=\frac{n}{v_0^d(t)}\chi(c,\beta), 
\quad \hbox{with} \quad 
\beta\equiv\frac{v_s}{v_0(t)}
\label{eq:scaling}
\ee
and again $\mathbf{c} = \mathbf{v}/v_0(t)$.
Note that we have assumed isotropy, $\chi$ depending on $c=|\mathbf{c}|$ and not on
the full vector $\mathbf{c}$, but this assumption can be easily relaxed.
Note also that equivalent expressions can of course be chosen, such as 
$\tilde\chi(v/v_s,\beta)$. 

If a state such as (\ref{eq:scaling}) holds, it represent a strong constraint
on the form of the velocity distribution. The corresponding dynamics can be partitioned
in a first rapid stage --that we do not attempt to describe-- 
where initial conditions matter, and a subsequent universal
relaxation towards stationarity, where only the distance to the steady state 
is relevant, through the dimensionless inverse typical velocity $\beta=v_s/v_0(t)$.

\subsection{Numerical simulations answer}
\label{ssec:numerics}

\begin{figure}[htb]
\begin{minipage}{0.48\linewidth}
\begin{center}
\includegraphics[angle=0,width=0.95\linewidth,clip]{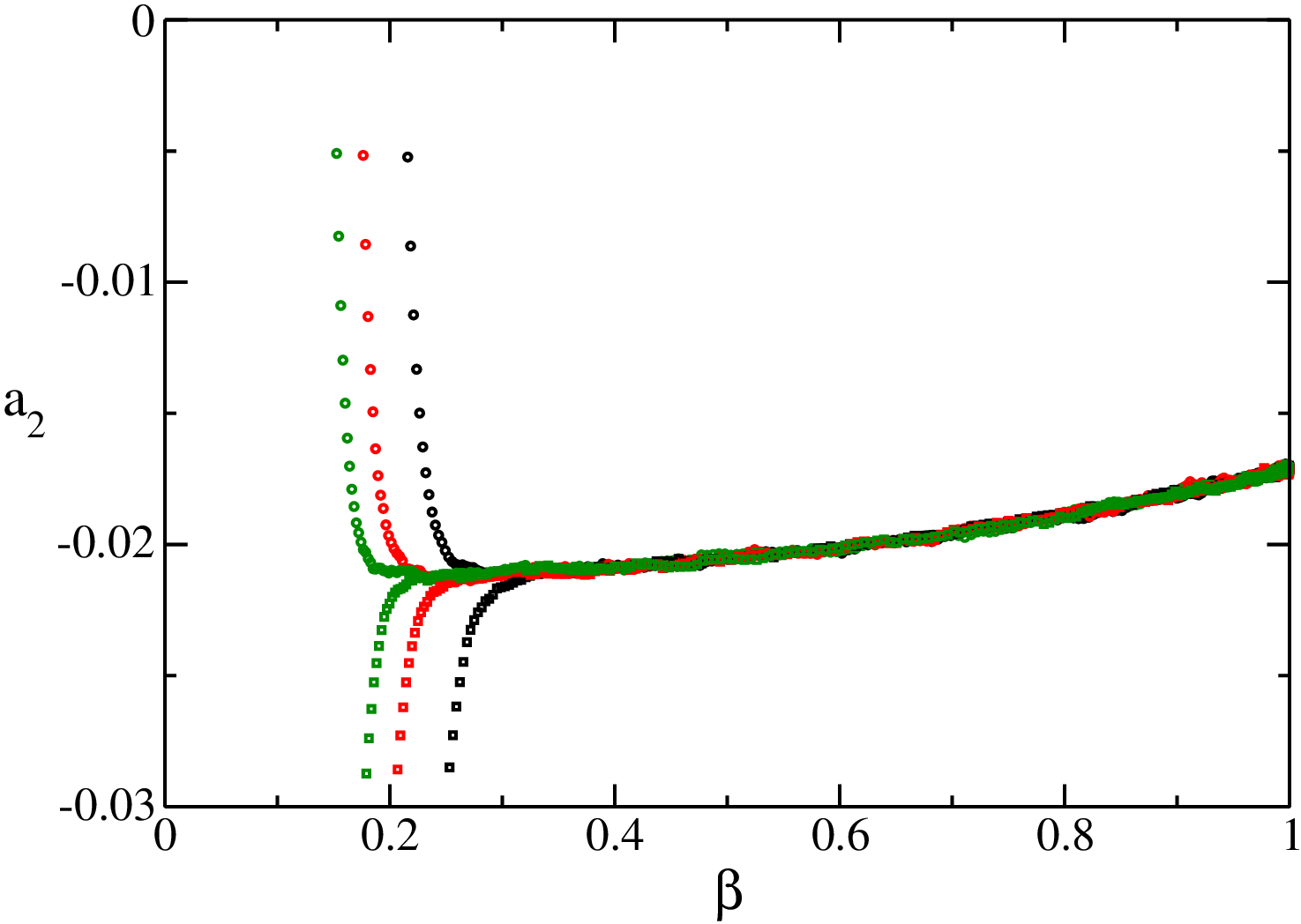}
\end{center}
\end{minipage}
\begin{minipage}{0.48\linewidth}
\begin{center}
\includegraphics[angle=0,width=0.95\linewidth,clip]{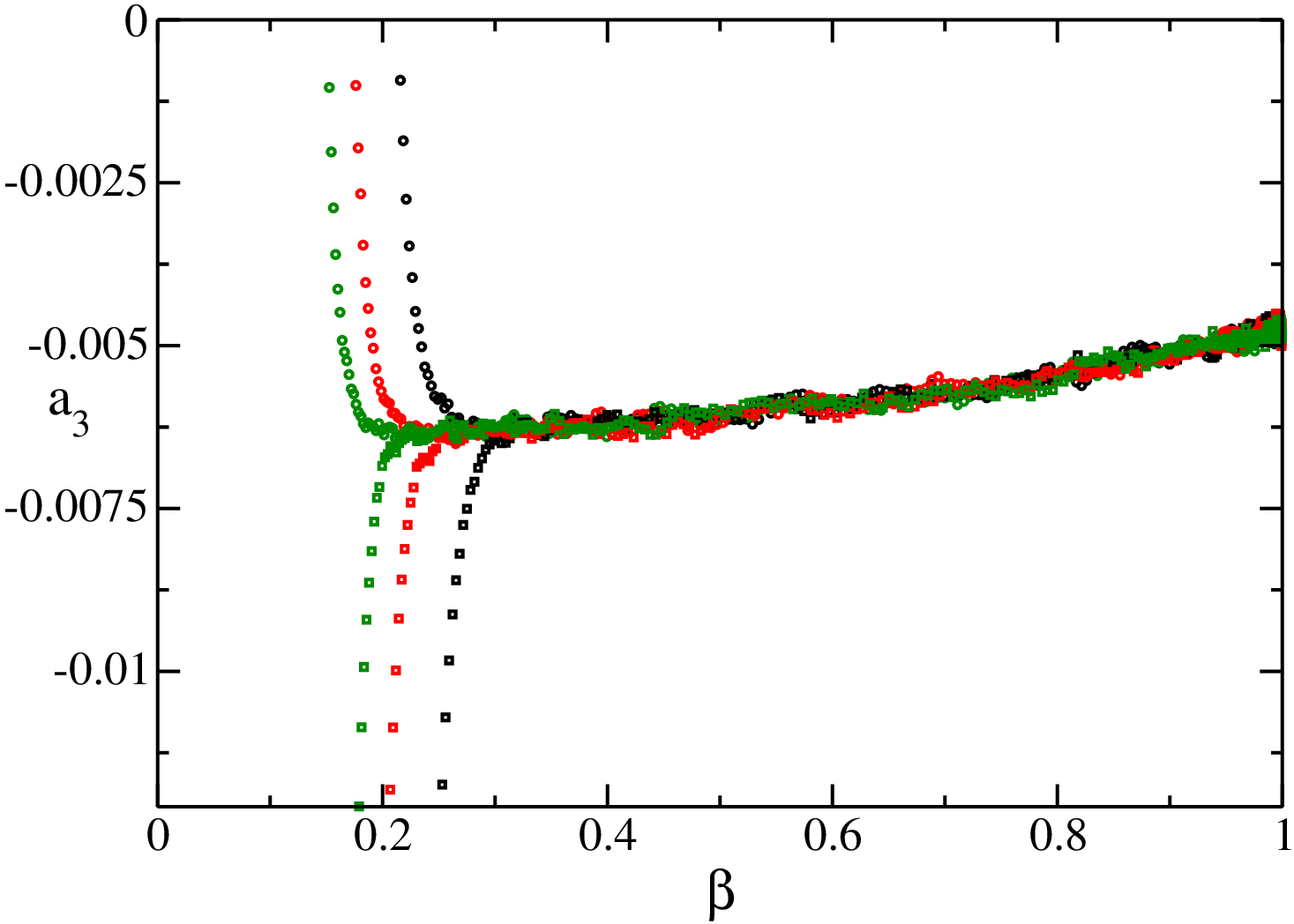}
\end{center}
\end{minipage}
\caption{Coefficients $a_2$ and $a_3$ from Monte Carlo,
for $\alpha=0.95$. We start with a Maxwellian (circles) or 
asymmetric  (squares) distribution, and three initial temperatures: $T_0/T_s= 22, 33, 44$.
Note the vertical scale.}
\label{fig1}
\end{figure}

\begin{figure}[htb]
\begin{minipage}{0.48\linewidth}
\begin{center}
\includegraphics[angle=0,width=0.95\linewidth,clip]{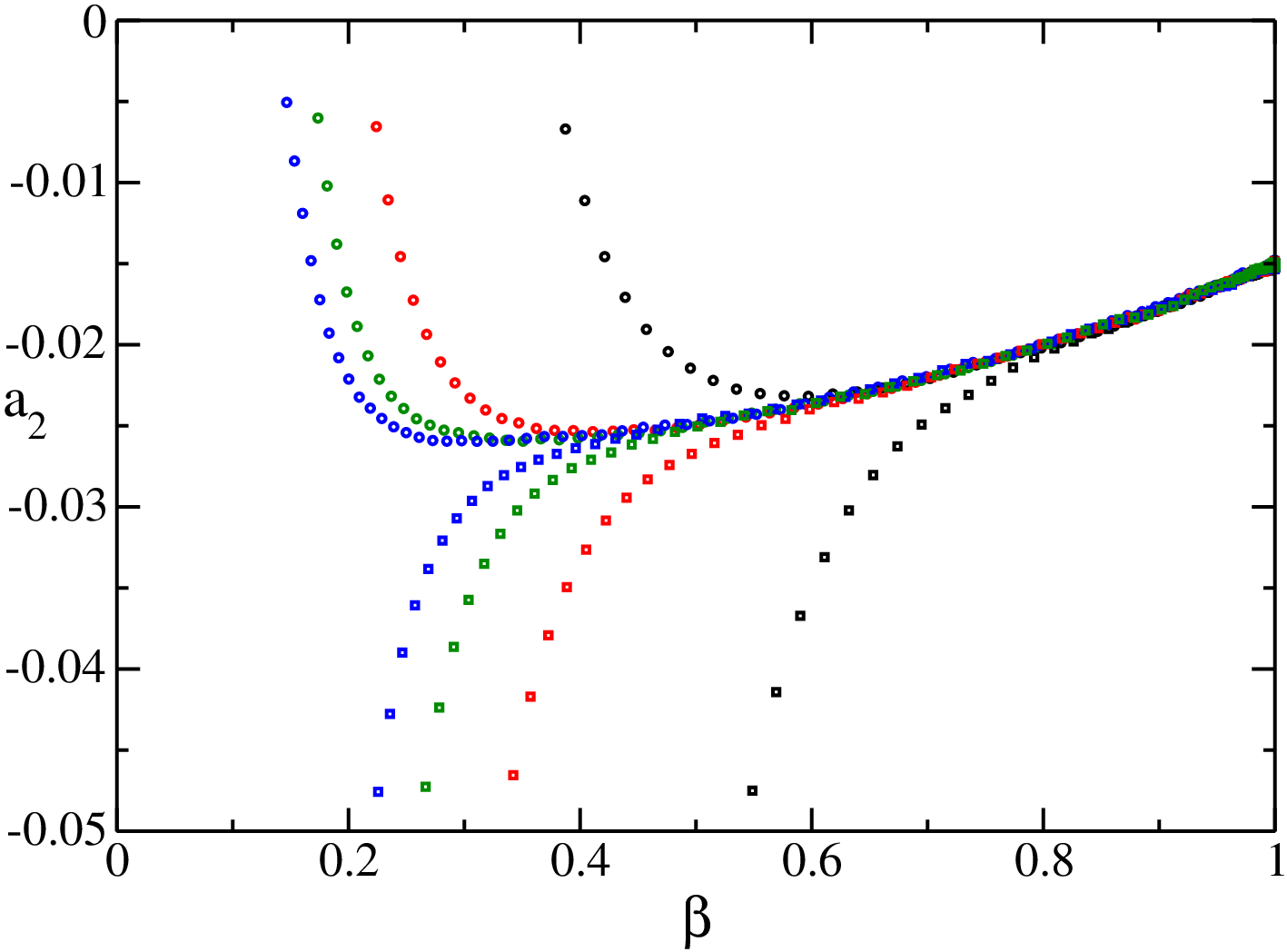}
\end{center}
\end{minipage}
\begin{minipage}{0.48\linewidth}
\begin{center}
\includegraphics[angle=0,width=0.95\linewidth,clip]{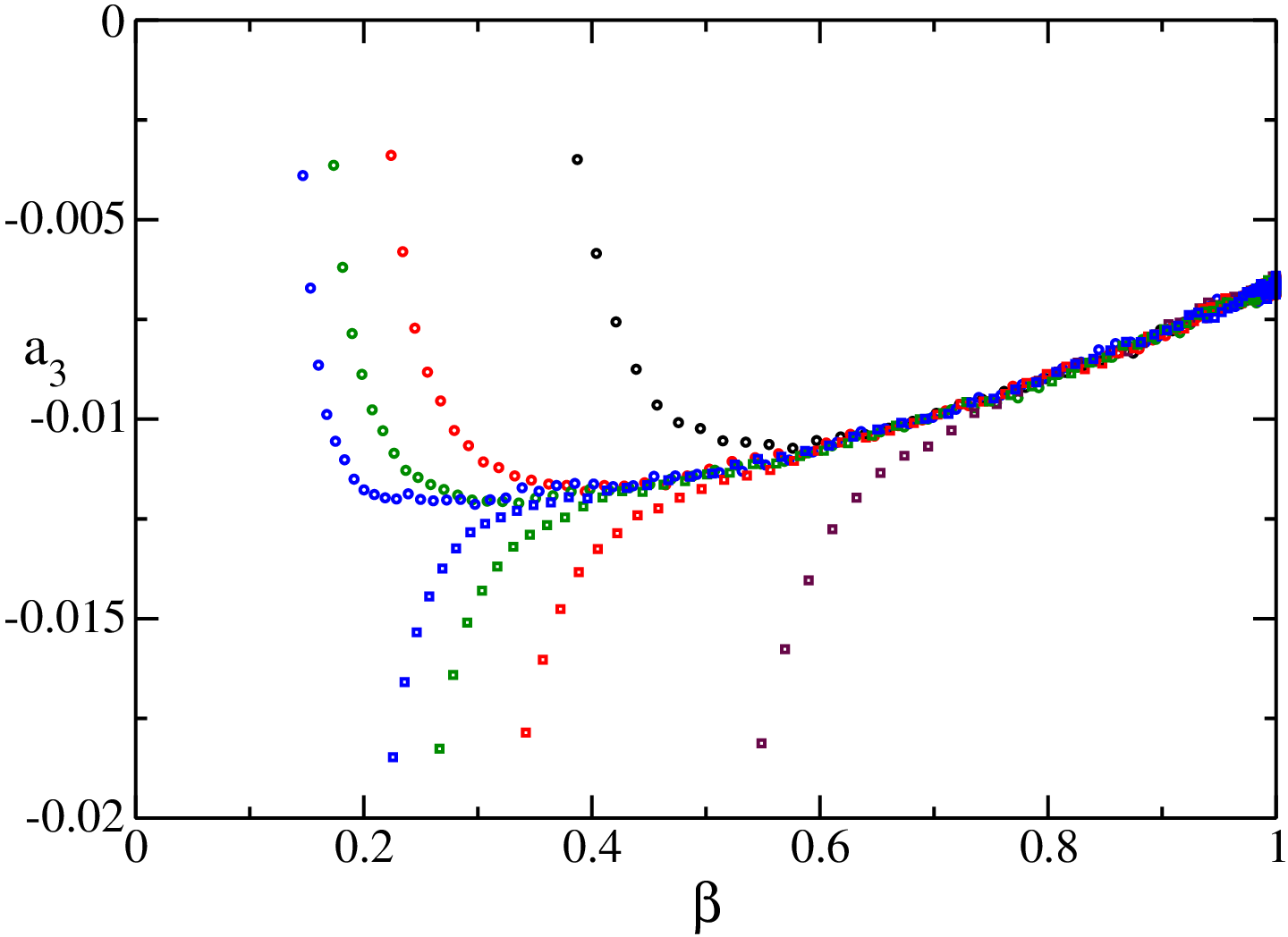}
\end{center}
\end{minipage}
\caption{Same as Fig \ref{fig1}, but for a more dissipative system 
with restitution coefficient $\alpha=0.8$. Here, the initial condition is either Gaussian
(circles, approaching the master curve from above), or flat (squares, see text, approaching the
master curve from below). In both cases, the initial temperature is 
$T_0/T_s= 7.3, 21.8, 36.4$ and 51.}
\label{fig2}
\end{figure}

To put the above scenario to the test, we have performed Direct 
Monte Carlo Simulations (DSMC) 
\cite{bird} of $N=1000$ hard disks ($d=2$) of unit mass and unit diameter, that 
collide inelastically with the collision rule given by equation 
(\ref{collisionRule}). The thermostat is implemented following previous investigations 
\cite{vnetp99}
and the results have been averaged over $10^5$ 
trajectories. For a given value of the 
inelasticity, we thus solve the time-dependent Boltzmann equation for different 
initial conditions and analyze whether, after some kinetic transient, 
all the time dependence of the 
distribution function goes through the dimensionless parameter $\beta$. As it 
is difficult to 
measure the complete distribution function with the desired accuracy, we have 
worked with the cumulants of the scaled distribution, 
$\chi(\mathbf{c},\beta)$. In terms of the velocity moments, 
$\langle v^l\rangle\equiv\frac{1}{n}\int d\mathbf{v}v^lf(\mathbf{v},t)$, we 
have measured the kurtosis of the distribution
\begin{equation}\label{kurtosis}
a_2=\frac{d}{d+2}\frac{\langle v^4\rangle}{\langle v^2\rangle^2}-1,  
\end{equation}
which is proportional to the fourth cumulant of $\chi(\mathbf{c},\beta)$ and 
the quantity
\begin{equation}
a_3=-\frac{d^2}{(d+2)(d+4)}\frac{\langle v^6\rangle}{\langle v^2\rangle^3}
+\frac{3d}{d+2}\frac{\langle v^4\rangle}{\langle v^2\rangle^2}-2, 
\end{equation}
which can be viewed as the reduced sixth cumulant. If our scaling is correct, 
we expect 
that the cumulants quickly collapse for different 
initial conditions, as a function of $\beta$. In Fig. \ref{fig1}, we have 
plotted $a_2$ and $a_3$ versus $\beta$ for $\alpha=0.95$. The initial 
conditions are either Maxwellian distributions with three different 
temperatures $T_0$, significantly above the steady state value $T_s$, or
asymmetric distributions made up of three possible velocities with different 
probabilities
\begin{eqnarray}
f(v_x,v_y,t=0)&=&\frac{3}{6}\delta\left(v_x+8D/3\right)\delta\left(v_y+8D/3\right)
+\frac{2}{6}\delta\left(v_x-4D/3\right)\delta\left(v_y-4D/3\right)\nonumber\\&+&\frac{1}{6}\delta\left(v_x-16D/3\right)\delta\left(v_y-16D/3\right).
\label{eq:asym}
\end{eqnarray}
Here, the parameter $D$ is chosen to match the initial desired temperature
(chosen the same as in the Gaussian initial condition).
All the quantities are measured every $250$ collisions, so that each 
four consecutive points in figure \ref{fig1} correspond to a time span of one 
collision per particle. It can be seen that, 
after some transient, memory of the initial condition is forgotten, so that the stationary 
distribution ($\beta=1$) is reached following a universal route. 
In figure \ref{fig1},
those data points associated to the Gaussian initial distribution approach
the scaling curve from above (circles) while those for the initial asymmetric
case (\ref{eq:asym}) approach the scaling curve from below (squares).
A very similar behavior can be seen in Fig \ref{fig2} for $\alpha=0.8$ and four different 
initial temperatures, again such that $T_0 \gg T_s$, which ensures that 
$\beta<1$ (we have also probed the regime $\beta>1$ obtained with $T_0 \ll T_s$,
where similar conclusions hold, see e.g. Fig. \ref{fig3} below). 
We have started either with a Maxwellian distribution, as above,
or with a distribution in which all the velocities have the same probability density in a 
square centered on $\mathbf{v}=\mathbf{0}$ (referred to as the ``flat'' case). 
We emphasize that the initial transient is fast: 
memory of the initial condition is lost after at most $3$ or $4$ collisions per 
particle, a phenomenon that cannot be appreciated from the figures. 

\begin{figure}[htb]
\begin{minipage}{0.48\linewidth}
\begin{center}
\includegraphics[angle=0,width=0.95\linewidth,clip]{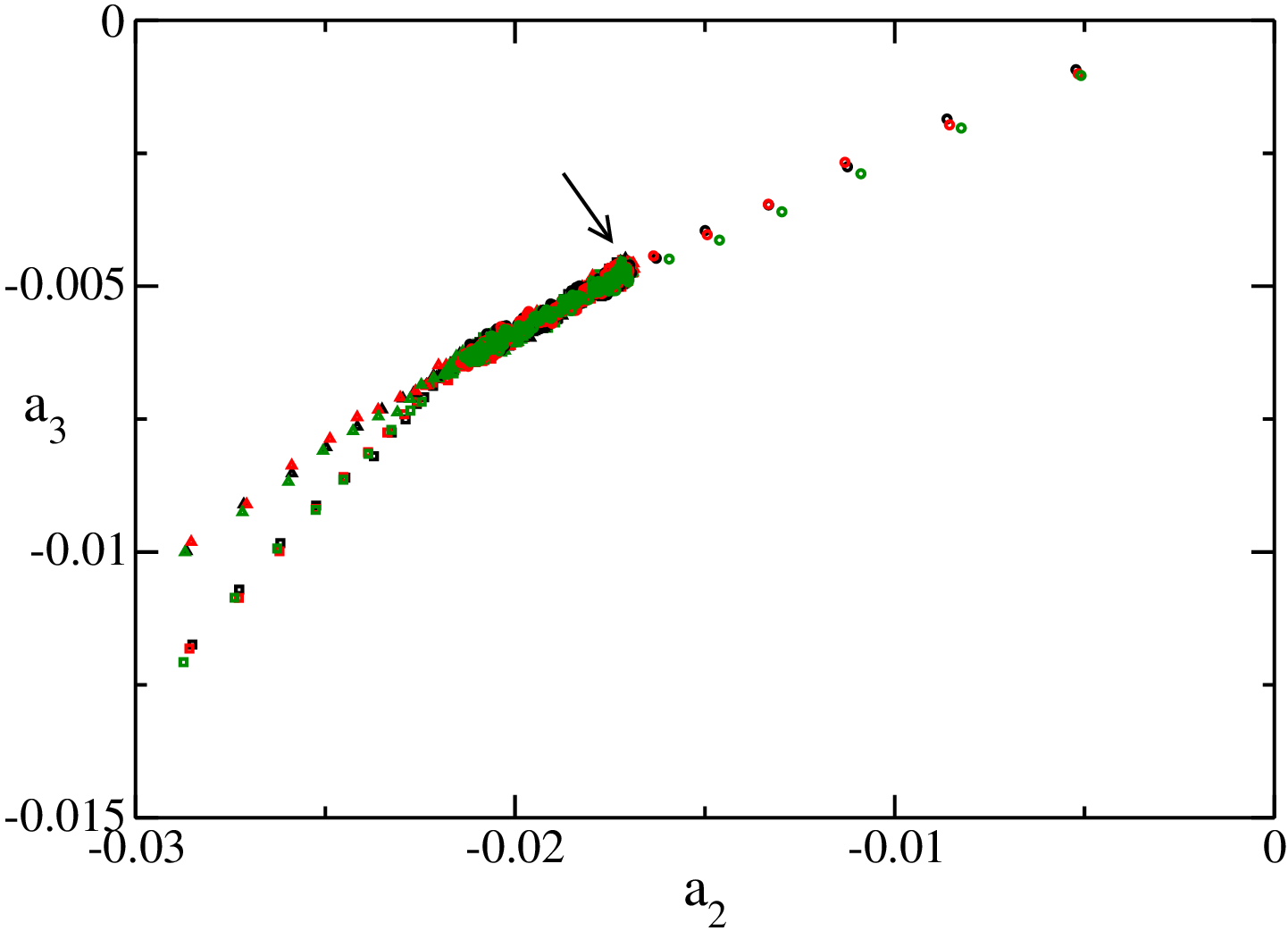}
\end{center}
\end{minipage}

\caption{Same data as in Fig. \ref{fig1}, 
where the coefficient $a_3$ is shown as a function of $a_2$.
In addition to Maxwellian and asymmetric, flat initial conditions
are also shown.
The arrow indicates the steady state values.}
\label{fig:ess}
\end{figure}

Borrowing ideas from the extended self-similarity technique \cite{B93},
we put to the test the possibility of an enhanced universality,
by plotting $a_3$ as a function of $a_2$, see Fig.  \ref{fig:ess}.
In doing so, it appears that the universal part of the $a_i$ versus $\beta$ 
curve seen in Fig. \ref{fig1} is not enhanced by the reparametrization 
$a_3(a_2)$: different initial conditions do not lead to a data 
collapse, beyond the interval $-0.22<a_2<-0.17$ that was already
evidenced in Fig. \ref{fig1}. However, a given functional form
(say Maxwellian) leads to a unique path in the $a_3$-$a_2$ plane,
which is already a non trivial point, and furthermore,
a plot like Fig. \ref{fig:ess} leads to a significantly reduced
scatter of points that Fig. \ref{fig1}.  It is therefore 
more amenable to chart out the universal regime sought for.
In these figures, the first measure reported
after the dynamics has acted 
on the initial conditions is for a time of $0.25$ collisions 
per particle for the Maxwellian distribution and of around $3$ collisions per particle for the flat and asymmetric distributions. 
The present results establish numerically the
existence of a universal non-trivial scaling regime, for which we now
seek analytical characterization.

\subsection{Analytical approach}
\label{ssec:analytical}

For the sake of analytical progress, it is convenient to change 
variables in the Boltzmann equation (\ref{ec.b}), from the set
$\{t,\mathbf{v}\}$, to $\{\beta,\mathbf{c}\}$. In these variables, the scaled 
distribution function fulfills
\begin{eqnarray}\label{ec.Scal}
[\mu(\beta)-\mu(1)\beta^3]\left\{\frac{\partial}{\partial\mathbf{c}_1}\cdot
[\mathbf{c}_1\chi(\mathbf{c}_1, \beta)]
+\beta\frac{\partial}{\partial\beta}\chi(\mathbf{c}_1,\beta )\right\}
\nonumber\\
=\int d\mathbf{c}_2\bar{T}_0(\mathbf{c}_1,\mathbf{c}_2)
\chi(\mathbf{c}_1,\beta)\chi(\mathbf{c}_2,\beta)+\frac{1}{2}\mu(1)\beta^3
\frac{\partial^2}{\partial\mathbf{c}_1^2}\chi(\mathbf{c}_1, \beta), 
\end{eqnarray}
where 
\be
\mu(\beta)=-\frac{1}{2d}\int d\mathbf{c}_1\int d\mathbf{c}_2(c_1^2+c_2^2)
\bar{T}_0(\mathbf{c}_1,\mathbf{c}_2)
\chi(\mathbf{c}_1, \beta)\chi(\mathbf{c}_2, \beta). 
\ee 
We note that  here and in contrast with Eq. (\ref{in.ec}), the 
equation is fully consistent as it appears as a change of variables where 
$\beta$ simply plays the role of time. 
Nevertheless, proving that 
for any ``reasonable'' initial condition, the system forgets the initial 
condition and reaches a universal state is a formidable 
task. For this reason we limit ourselves to the simplified problem of 
deriving an approximate expression for this distribution function. 
As in the stationary state, the distribution will be worked out in 
the first Sonine approximation
\begin{equation}\label{sonExp}
\chi(c,\beta)\approx\chi_M(c)[1+a_2(\beta)S_2(c^2)],
\end{equation}
where the kurtosis, $a_2$, has been defined in (\ref{kurtosis}) and, by 
definition, we have
\be
\int d\mathbf{c}\chi(c,\beta)=1, 
\qquad \int d\mathbf{c}\mathbf{c}\chi(c,\beta)=\mathbf{0}, 
\qquad \int d\mathbf{c}c^2\chi(c,\beta)=\frac{d}{2}. 
\ee
In expansion (\ref{sonExp}), we neglect contributions in $a_3$
and higher order. This is justified as long as the inelasticity 
is not too strong, and is backed up here by the fact 
that $|a_3|<a_2$, as can be seen in 
Figs. \ref{fig1} and \ref{fig2}. Of course, inclusion of 
higher order terms in (\ref{sonExp}) would improve the accuracy
of the subsequent calculation.

Inserting (\ref{sonExp}) into the Boltzmann equation (\ref{ec.Scal}), 
taking the fourth velocity moment while neglecting nonlinear terms in $a_2$, 
we obtain the 
following evolution equation for the cumulant $a_2$
\be\label{ec.a2}
\frac{1}{4}\beta(1-\beta^3)\frac{d}{d\beta}a_2(\beta)
=(1-B-\beta^3)a_2(\beta)+Ba_2^s, 
\ee
where the parameter $B$ depends on dissipation and space dimension according to
\be
B=\frac{73+8d(7-3\alpha)+15\alpha[2\alpha(1-\alpha)-7]}
{16(1-\alpha)(3+2d+2\alpha^2)
+a_2^s[85+d(30\alpha-62)+3\alpha(10\alpha(1-\alpha)-39)]}. 
\label{eq:B}
\ee

Eq. (\ref{ec.a2}) is an inhomogeneous linear differential equation that can 
be integrated. It exhibits two singular points at $\beta=0$, $\beta=1$, 
and we start with the interval $[0,1]$. 
In this case the general solution of the associated homogeneous equation reads
\be
a_2^H(\beta)=K\frac{(1-\beta^3)^{\frac{4}{3}B}}{\beta^{4(B-1)}}
\label{eq:part}
\ee
and a particular solution can be obtained by variations of parameters.
The general solution will then be the sum of these two contributions.
The ensuing $a_2$ depends on the initial conditions through 
$K$, and since our purpose here is to extract the universal
behaviour of $a_2$ as a function of $\beta$, we note that 
the contribution (\ref{eq:part}) fades rapidly as $\beta$ approaches 
unity (as $(1-\beta^3)^{4B/3}$ where $B$ can be large ; note that 
it diverges in the elastic limit $\alpha \to 1$). 
The universal behaviour is consequently encoded is the particular solution, 
and we finally have 
\be
a_2(\beta)=a_2^s\left[1+\frac{1-\beta^3}{B-1}
\,_2F_1\left(-\frac{1}{3},1;\frac{4B-1}{3};\beta^3\right)\right], 
\qquad 0<\beta<1. 
\label{eq:betapetit}
\ee
where $\,_2F_1$ is the hyper-geometric function \cite{arnoldVasilii}. 
This expression is well behaved in all the interval $[0,1]$. 
An analogous 
analysis can be performed for $\beta>1$. Following similar lines, we 
identify the universal solution to be
\be
a_2(\beta)=-\frac{4Ba_2^s}{7\beta^3(1-1/\beta^3)^{\frac{4B}{3}}}
\,_2F_1\left(\frac{7}{3},1+\frac{4B}{3};\frac{10}{3};\frac{1}{\beta^3}\right). 
\label{eq:betagrand}
\ee
Clearly, the same technical procedure can be applied to the 
higher order cumulants. For the sake
of simplicity, we restrict to the function $a_2(\beta)$, that we wish to
test against simulation results.

\begin{figure}
\begin{minipage}{0.48\linewidth}
\begin{center}
\includegraphics[angle=0,width=0.95\linewidth]{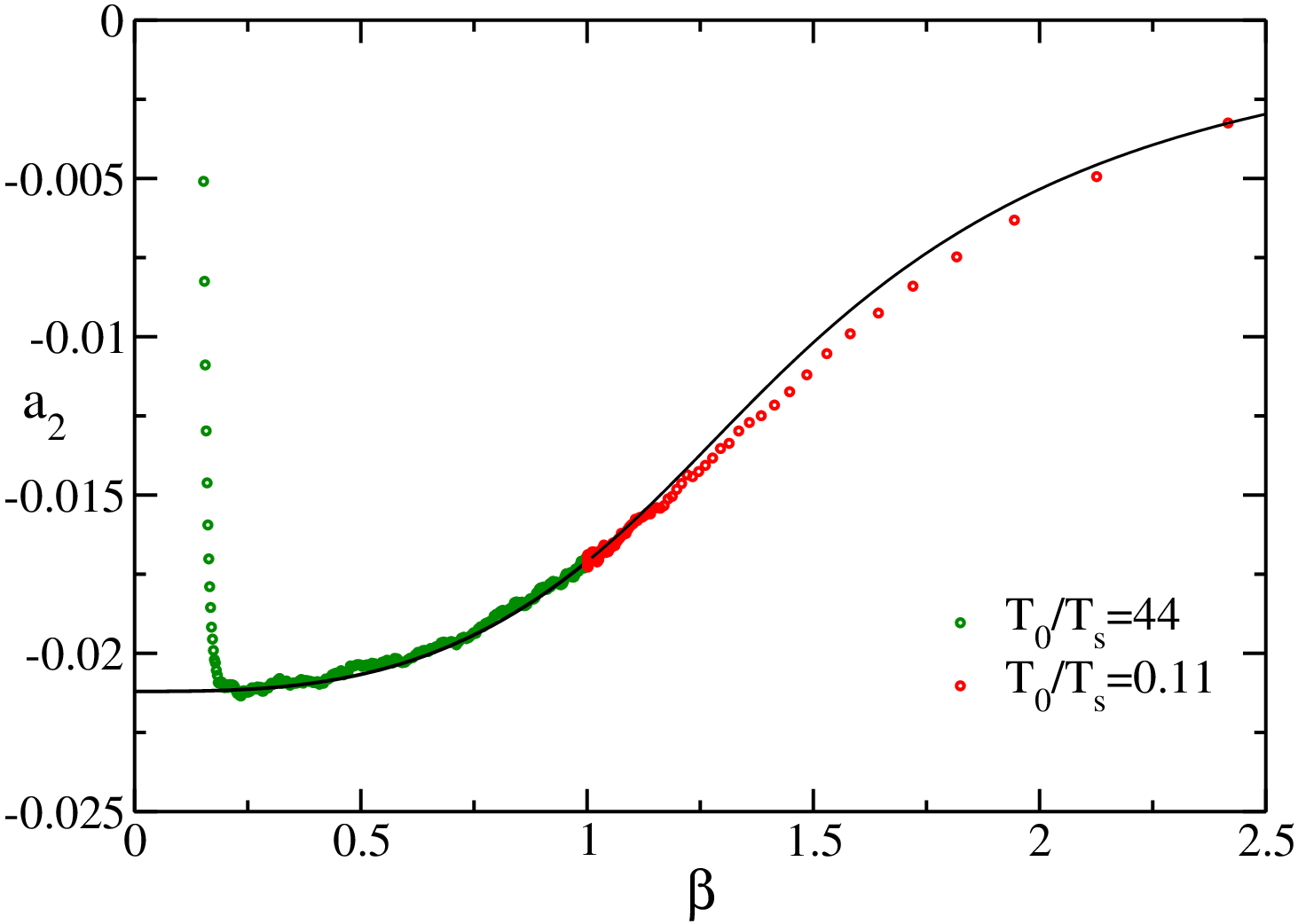}
\end{center}
\end{minipage}
\begin{minipage}{0.48\linewidth}
\begin{center}
\includegraphics[angle=0,width=0.95\linewidth]{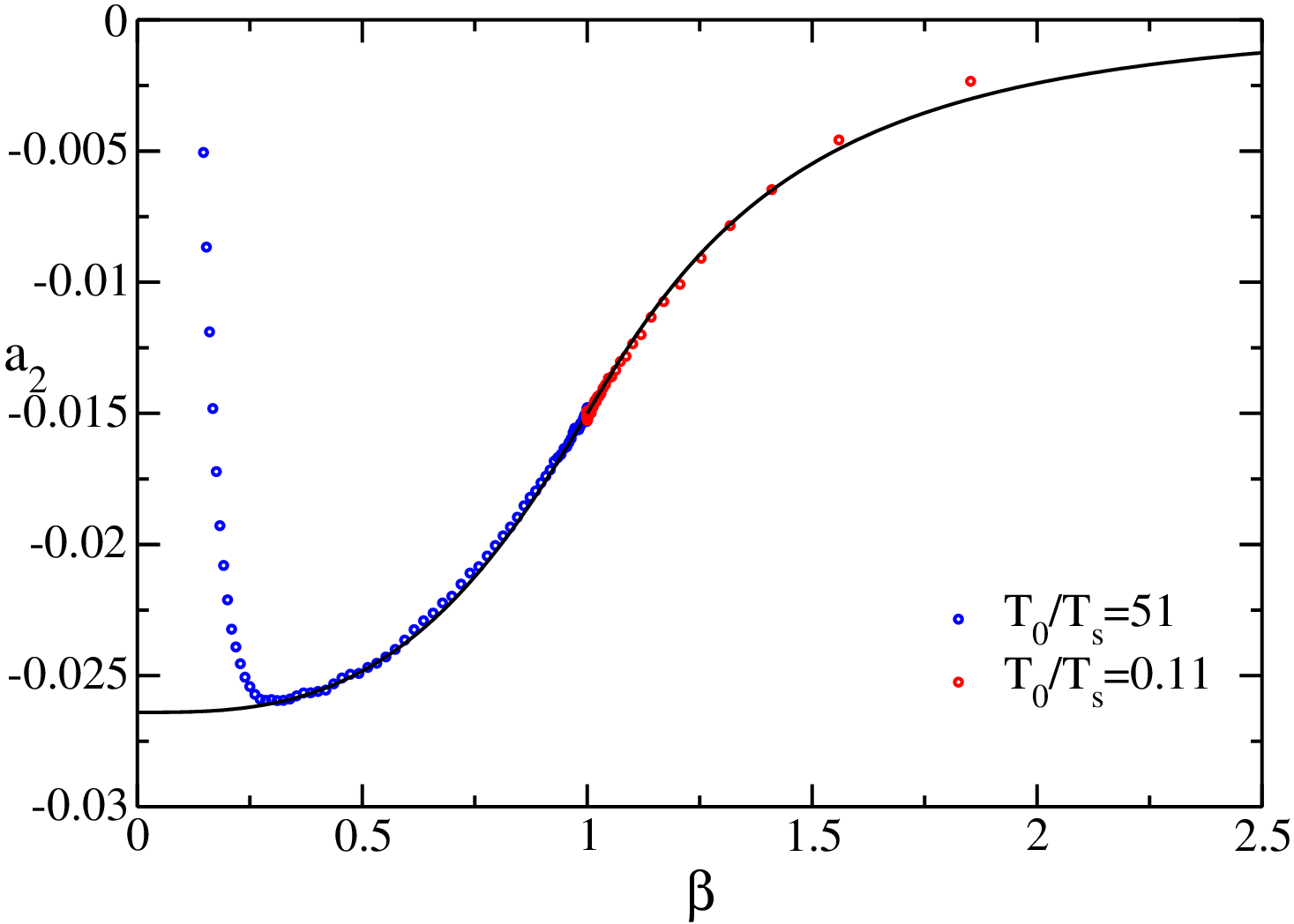}
\end{center}
\end{minipage}
\caption{Kurtosis as a function of $\beta$ (reduced inverse typical velocity), for a system with $\alpha=0.95$ (left) and $\alpha=0.80$ (right). The points are simulation results starting from a Maxwellian distribution and the solid line is the theoretical prediction, given by Eq. (\ref{eq:betapetit}) for $\beta<1$ and Eq. (\ref{eq:betagrand}) for $\beta>1$, see text. The two values of the initial temperature mentioned for each graph
are used to generate separately the $\beta<1$ branch (associated to large $T_0$) and the $\beta>1$
branch (obtained for small $T_0$). The steady state values correspond to $\beta=1$.}
\label{fig3}
\end{figure}

In order to compare the above theoretical predictions to the simulation data,
attention should be payed to the fact that analytical computation of 
velocity moments or cumulants is plagued by non-linear effects that have been discussed
in the literature \cite{MS00,C03}. This results in some error in the
calculation of the steady state value $a_2^s$, and we can also expect 
$B$ to suffer from a similar inaccuracy, that may be of the order
of 10 or 20\%. To circumvent this (somewhat minor) drawback, we take $a_2^s$ appearing
in Eqs. (\ref{eq:betapetit}) and (\ref{eq:betagrand}) from the Monte Carlo
simulations, and we adjust $B$ to match the measured function $a_2(\beta)$. 
This procedure provides us with Fig. \ref{fig3}, where the agreement
between the functional forms (\ref{eq:betapetit}) and (\ref{eq:betagrand})
with Monte Carlo is excellent. 
It is of course important to check {\it a posteriori}
that $a_2^s$ and $B$ thereby obtained are close enough to our predictions.
The precise values are reported below:
\begin{equation}
\begin{array}{l || c | c}
 \null                                   & \alpha=0.95     & \alpha=0.8  \\ 
 \hline\hline
 a_2^s \hbox{ from DSMC }                &    -0.0171          &   -0.0150    \\
 a_2^s \hbox{ from Eq. (\ref{eq:a2s}) }      &        -0.0157         &   -0.0135 \\
 \hline
 B \hbox{ from DSMC }                &        5.17         &   2.32     \\
 B \hbox{ from Eq. (\ref{eq:B}) }      &       6.43         &   2.61 
\end{array}
\end{equation}

\section{Conclusion and perspectives}
\label{sec:concl}

To summarize, we have studied the dynamics of a system of  
inelastic hard spherical grains, heated homogeneously by a stochastic thermostat. 
We have restricted the analysis to low density systems, amenable to 
a Boltzmann equation treatment. We have found that generically, after a kinetic transient, 
the system evolves into a scaling solution that no longer depends on initial conditions,
before the steady state is finally reached. The relevant scaling form is not of the
single parameter family as is the case in the homogeneous cooling state, 
but involves 2 parameters. The velocity distribution 
function was calculated in the so-called first Sonine approximation, 
which provides a very good 
agreement with the Monte Carlo simulations. 

At this point several questions arise: What is the counterpart of the 
universal state brought to the fore at two-particle level, or even $N$-particle ?
What is the effect of density, and of a change in the driving
mechanism ? 
Do the hydrodynamic relations (worked out say at Chapman Enskog level) 
depend on the structure of this state? 
The complete answer to all these interrogations requires further studies,
but we expect that similar scaling forms should occur at higher densities,
and with different thermostats as long as a steady state can be reached.
In this respect, the Gaussian thermostat is presumably singular,
since it can be mapped onto the free cooling case. 
The questions pertaining to hydrodynamics are more subtle;
How the universal $\beta$-scaling behaviour discussed in the present work
impinges, as a reference state,
on transport properties, should be explored.

\section{Acknowledgments}
This research was supported by the Ministerio de Educaci\'{o}n y
Ciencia (Spain) through Grant No. FIS2008-01339 (partially financed
by FEDER funds).

\end{document}